\pdfoutput=1

\documentclass[12pt,a4paper]{article}

\usepackage[colorinlistoftodos]{todonotes}

\usepackage{ifthen} 
\newboolean{pdflatex}
\setboolean{pdflatex}{true} 

\newboolean{articletitles}
\setboolean{articletitles}{true} 

\newboolean{uprightparticles}
\setboolean{uprightparticles}{false} 

\newboolean{inbibliography}
\setboolean{inbibliography}{false} 

\def\paperauthors{LHCb collaboration} 
\def\paperasciititle{Measurement of the Omegac baryon lifetime} 
\def\papertitle{Measurement of the $\Omegac$ baryon lifetime} 
\def\paperkeywords{{High Energy Physics}, {LHCb}} 
\def\papercopyright{\the\year\ CERN for the benefit of the LHCb collaboration}
\def\paperlicence{CC-BY-4.0}
\def\paperlicenceurl{https://creativecommons.org/licenses/by/4.0/}

\usepackage[top=1in, bottom=1.25in, left=1in, right=1in]{geometry}

\usepackage{microtype}
\usepackage{lineno}  
\usepackage{xspace} 
\usepackage{caption} 

\usepackage{graphicx}  
\usepackage{color}
\usepackage{colortbl}
\graphicspath{{./figs/}} 

\usepackage{amsmath} 
\usepackage{amssymb}
\usepackage{amsfonts}
\usepackage{upgreek} 

\newcommand*\patchAmsMathEnvironmentForLineno[1]{%
\expandafter\let\csname old#1\expandafter\endcsname\csname #1\endcsname
\expandafter\let\csname oldend#1\expandafter\endcsname\csname
end#1\endcsname
 \renewenvironment{#1}%
   {\linenomath\csname old#1\endcsname}%
   {\csname oldend#1\endcsname\endlinenomath}%
}
\newcommand*\patchBothAmsMathEnvironmentsForLineno[1]{%
  \patchAmsMathEnvironmentForLineno{#1}%
  \patchAmsMathEnvironmentForLineno{#1*}%
}
\AtBeginDocument{%
\patchBothAmsMathEnvironmentsForLineno{equation}%
\patchBothAmsMathEnvironmentsForLineno{align}%
\patchBothAmsMathEnvironmentsForLineno{flalign}%
\patchBothAmsMathEnvironmentsForLineno{alignat}%
\patchBothAmsMathEnvironmentsForLineno{gather}%
\patchBothAmsMathEnvironmentsForLineno{multline}%
\patchBothAmsMathEnvironmentsForLineno{eqnarray}%
}

\usepackage{hyperxmp}

\usepackage[pdftex,
            pdfauthor={\paperauthors},
            pdftitle={\paperasciititle},
            pdfkeywords={\paperkeywords},
            pdfcopyright={Copyright (C) \papercopyright},
            pdflicenseurl={\paperlicenceurl}]{hyperref}

\usepackage[all]{hypcap} 

\usepackage{xspace} 
\usepackage{upgreek}

\def\lhcb {\mbox{LHCb}\xspace}

\def\MagUp {\mbox{\em Mag\kern -0.05em Up}\xspace}

\ifthenelse{\boolean{uprightparticles}}%
{

 \def\Pmu         {\ensuremath{\upmu}\xspace}                 
 \def\Pnu         {\ensuremath{\upnu}\xspace}                 
                  
 \def\Ppi         {\ensuremath{\uppi}\xspace}

 \def\Ptau        {\ensuremath{\uptau}\xspace}

 \def\PDelta      {\ensuremath{\Delta}\xspace}                 
 \def\PXi      {\ensuremath{\Xi}\xspace}                 
 \def\PLambda      {\ensuremath{\Lambda}\xspace}                 
 \def\PSigma      {\ensuremath{\Sigma}\xspace}                 
 \def\POmega      {\ensuremath{\Omega}\xspace}                 
 \def\PUpsilon      {\ensuremath{\Upsilon}\xspace}                 
 
 \def\PB      {\ensuremath{\mathrm{B}}\xspace}                 
                  
 \def\PD      {\ensuremath{\mathrm{D}}\xspace}

 \def\PK      {\ensuremath{\mathrm{K}}\xspace}

 \def\Pb      {\ensuremath{\mathrm{b}}\xspace}                 
 \def\Pc      {\ensuremath{\mathrm{c}}\xspace}

 \def\Pi      {\ensuremath{\mathrm{i}}\xspace}

 \def\Ps      {\ensuremath{\mathrm{s}}\xspace}

}
{

 \def\Pmu         {\ensuremath{\mu}\xspace}                 
 \def\Pnu         {\ensuremath{\nu}\xspace}                 
                  
 \def\Ppi         {\ensuremath{\pi}\xspace}

 \def\Ptau        {\ensuremath{\tau}\xspace}

 \mathchardef\PDelta="7101
 \mathchardef\PXi="7104
 \mathchardef\PLambda="7103
 \mathchardef\PSigma="7106
 \mathchardef\POmega="710A
 \mathchardef\PUpsilon="7107
                  
 \def\PB      {\ensuremath{B}\xspace}                 
                  
 \def\PD      {\ensuremath{D}\xspace}

 \def\PK      {\ensuremath{K}\xspace}

 \def\Pb      {\ensuremath{b}\xspace}                 
 \def\Pc      {\ensuremath{c}\xspace}

 \def\Pi      {\ensuremath{i}\xspace}

 \def\Ps      {\ensuremath{s}\xspace}

}

\makeatletter
\ifcase \@ptsize \relax
  \newcommand{\miniscule}{\@setfontsize\miniscule{4}{5}}
\or
  \newcommand{\miniscule}{\@setfontsize\miniscule{5}{6}}
\or
  \newcommand{\miniscule}{\@setfontsize\miniscule{5}{6}}
\fi
\makeatother

\DeclareRobustCommand{\optbar}[1]{\shortstack{{\miniscule (\rule[.5ex]{1.25em}{.18mm})}
  \\ [-.7ex] $#1$}}

\def\mun        {{\ensuremath{\Pmu^-}}\xspace} 

\def\taum       {{\ensuremath{\Ptau^-}}\xspace}

\def\neu        {{\ensuremath{\Pnu}}\xspace}
\def\neub       {{\ensuremath{\overline{\Pnu}}}\xspace}

\def\neumb      {{\ensuremath{\neub_\mu}}\xspace}
\def\neut       {{\ensuremath{\neu_\tau}}\xspace}
\def\neutb      {{\ensuremath{\neub_\tau}}\xspace}

\def\squark    {{\ensuremath{\Ps}}\xspace}

\def\cquark    {{\ensuremath{\Pc}}\xspace}

\def\bquark    {{\ensuremath{\Pb}}\xspace}

\def\pion   {{\ensuremath{\Ppi}}\xspace}

\def\pip    {{\ensuremath{\pion^+}}\xspace}
\def\pim    {{\ensuremath{\pion^-}}\xspace}

\def\kaon    {{\ensuremath{\PK}}\xspace}
  \def\Kbar    {{\kern 0.2em\overline{\kern -0.2em \PK}{}}\xspace}

\def\KorKbar    {\kern 0.18em\optbar{\kern -0.18em K}{}\xspace}

\def\Kp      {{\ensuremath{\kaon^+}}\xspace}
\def\Km      {{\ensuremath{\kaon^-}}\xspace}

  \def\Dbar    {{\kern 0.2em\overline{\kern -0.2em \PD}{}}\xspace}
\def\D       {{\ensuremath{\PD}}\xspace}

\def\DorDbar    {\kern 0.18em\optbar{\kern -0.18em D}{}\xspace}
\def\Dz      {{\ensuremath{\D^0}}\xspace}
\def\Dzb     {{\ensuremath{\Dbar{}^0}}\xspace}
\def\Dp      {{\ensuremath{\D^+}}\xspace}
\def\Dm      {{\ensuremath{\D^-}}\xspace}

\def\Ds      {{\ensuremath{\D^+_\squark}}\xspace}
\def\Dsp     {{\ensuremath{\D^+_\squark}}\xspace}
\def\Dsm     {{\ensuremath{\D^-_\squark}}\xspace}

\def\B       {{\ensuremath{\PB}}\xspace}
\def\Bbar    {{\ensuremath{\kern 0.18em\overline{\kern -0.18em \PB}{}}}\xspace}

\def\BorBbar    {\kern 0.18em\optbar{\kern -0.18em B}{}\xspace}

\def\Bub     {{\ensuremath{\B^-}}\xspace}

\def\Bm      {{\ensuremath{\Bub}}\xspace}

\def\Bs      {{\ensuremath{\B^0_\squark}}\xspace}

  \def\Y#1S{\ensuremath{\PUpsilon{(#1S)}}\xspace}

\def\Xires       {{\ensuremath{\PXi}}\xspace}

\def\Lz          {{\ensuremath{\PLambda}}\xspace}
\def\Lbar        {{\ensuremath{\kern 0.1em\overline{\kern -0.1em\PLambda}}}\xspace}
\def\LorLbar    {\kern 0.18em\optbar{\kern -0.18em \PLambda}{}\xspace}

\def\Omegares    {{\ensuremath{\POmega}}\xspace}

\def\Lc      {{\ensuremath{\Lz^+_\cquark}}\xspace}

\def\Xicz    {{\ensuremath{\Xires^0_\cquark}}\xspace}
\def\Xicp    {{\ensuremath{\Xires^+_\cquark}}\xspace}

\def\Omegac    {{\ensuremath{\Omegares^0_\cquark}}\xspace}

\def\Omegab    {{\ensuremath{\Omegares^-_\bquark}}\xspace}

\def\to                 {\ensuremath{\rightarrow}\xspace}

\def\AT#1     {\ensuremath{A_{\mathrm{T}}^{#1}}\xspace}           

\def\C#1      {\ensuremath{\mathcal{C}_{#1}}\xspace}                       
\def\Cp#1     {\ensuremath{\mathcal{C}_{#1}^{'}}\xspace}                    
\def\Ceff#1   {\ensuremath{\mathcal{C}_{#1}^{\mathrm{(eff)}}}\xspace}        
\def\Cpeff#1  {\ensuremath{\mathcal{C}_{#1}^{'\mathrm{(eff)}}}\xspace}       
\def\Ope#1    {\ensuremath{\mathcal{O}_{#1}}\xspace}                       
\def\Opep#1   {\ensuremath{\mathcal{O}_{#1}^{'}}\xspace}                    

\newcommand{\tev}{\ifthenelse{\boolean{inbibliography}}{\ensuremath{~T\kern -0.05em eV}}{\ensuremath{\mathrm{\,Te\kern -0.1em V}}}\xspace}
\newcommand{\gev}{\ensuremath{\mathrm{\,Ge\kern -0.1em V}}\xspace}
\newcommand{\mev}{\ensuremath{\mathrm{\,Me\kern -0.1em V}}\xspace}
\newcommand{\kev}{\ensuremath{\mathrm{\,ke\kern -0.1em V}}\xspace}
\newcommand{\ev}{\ensuremath{\mathrm{\,e\kern -0.1em V}}\xspace}
\newcommand{\gevc}{\ensuremath{{\mathrm{\,Ge\kern -0.1em V\!/}c}}\xspace}
\newcommand{\mevc}{\ensuremath{{\mathrm{\,Me\kern -0.1em V\!/}c}}\xspace}
\newcommand{\gevcc}{\ensuremath{{\mathrm{\,Ge\kern -0.1em V\!/}c^2}}\xspace}
\newcommand{\gevgevcccc}{\ensuremath{{\mathrm{\,Ge\kern -0.1em V^2\!/}c^4}}\xspace}
\newcommand{\mevcc}{\ensuremath{{\mathrm{\,Me\kern -0.1em V\!/}c^2}}\xspace}

\def\mum  {\ensuremath{{\,\upmu\mathrm{m}}}\xspace}

\def\invfb   {\ensuremath{\mbox{\,fb}^{-1}}\xspace}

\newcommand{\stat}{\ensuremath{\mathrm{\,(stat)}}\xspace}

\newcommand{\chisq}{\ensuremath{\chi^2}\xspace}

\newcommand{\chisqip}{\ensuremath{\chi^2_{\text{IP}}}\xspace}

\def\gsim{{~\raise.15em\hbox{$>$}\kern-.85em
          \lower.35em\hbox{$\sim$}~}\xspace}
\def\lsim{{~\raise.15em\hbox{$<$}\kern-.85em
          \lower.35em\hbox{$\sim$}~}\xspace}

\def\sPlot{\mbox{\em sPlot}\xspace}

\def\ptot       {\mbox{$p$}\xspace}
\def\pt         {\mbox{$p_{\mathrm{ T}}$}\xspace}

\def\evtgen     {\mbox{\textsc{EvtGen}}\xspace}

\def\geant      {\mbox{\textsc{Geant4}}\xspace}

\def\photos     {\mbox{\textsc{Photos}}\xspace}

\def\pythia     {\mbox{\textsc{Pythia}}\xspace}

\def\tell1  {TELL1\xspace}
\def\ukl1   {UKL1\xspace}

\usepackage{cite} 
\usepackage{mciteplus}

\usepackage{longtable} 

\begin{document}

\renewcommand{\thefootnote}{\fnsymbol{footnote}}
\setcounter{footnote}{1}


\begin{titlepage}
\pagenumbering{roman}

\vspace*{-1.5cm}
\centerline{\large EUROPEAN ORGANIZATION FOR NUCLEAR RESEARCH (CERN)}
\vspace*{1.5cm}
\noindent
\begin{tabular*}{\linewidth}{lc@{\extracolsep{\fill}}r@{\extracolsep{0pt}}}
\ifthenelse{\boolean{pdflatex}}
{\vspace*{-1.5cm}\mbox{\!\!\!\includegraphics[width=.14\textwidth]{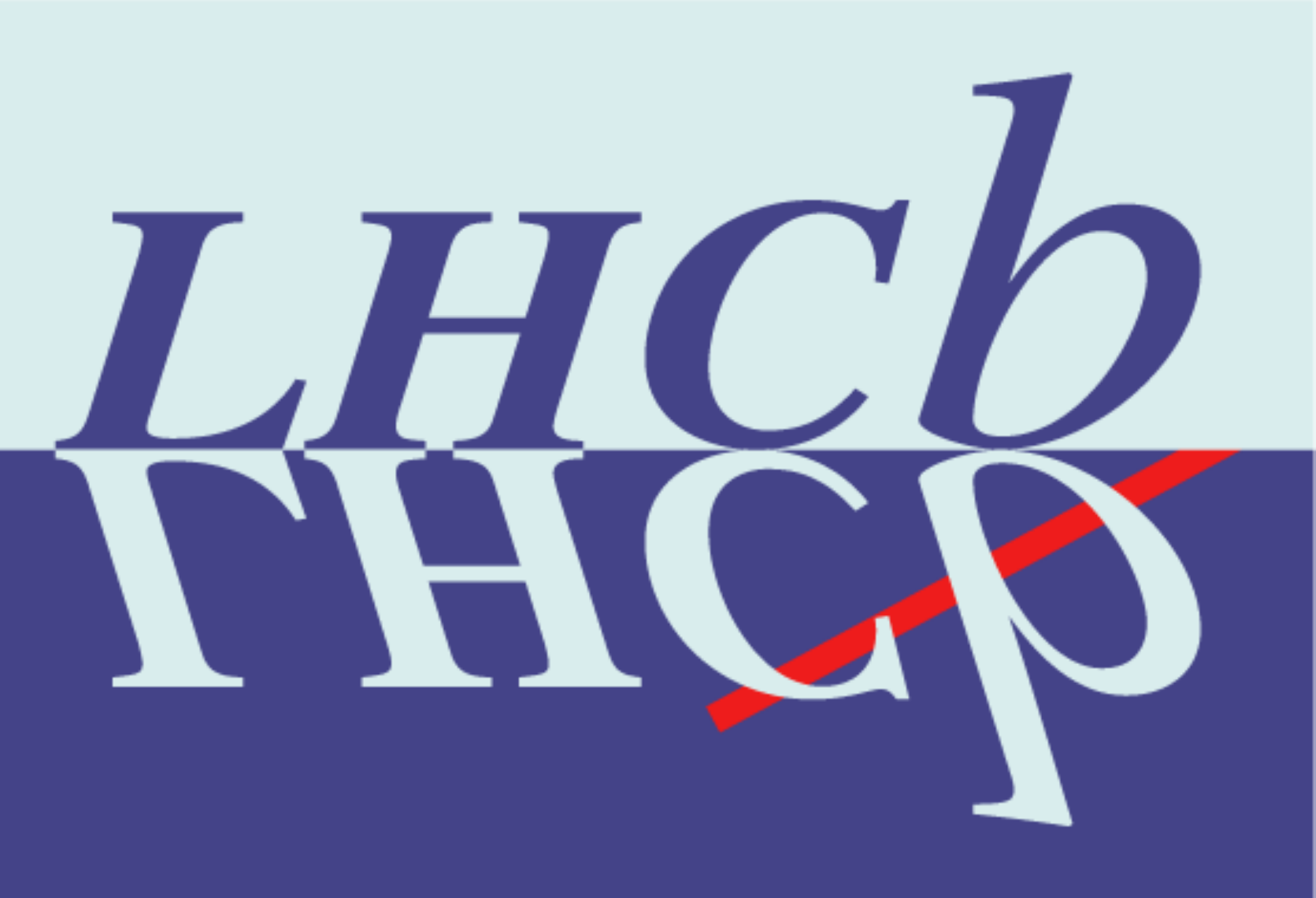}} & &}%
{\vspace*{-1.2cm}\mbox{\!\!\!\includegraphics[width=.12\textwidth]{lhcb-logo.eps}} & &}%
\\
 & & CERN-EP-2018-175 \\  
 & & LHCb-PAPER-2018-028 \\  
 & & July 5, 2018 \\ 
 & & \\
\end{tabular*}

\vspace*{4.0cm}

{\normalfont\bfseries\boldmath\huge
\begin{center}
  \papertitle 
\end{center}
}

\vspace*{2.0cm}

\begin{center}
\paperauthors\footnote{Authors are listed at the end of this paper.}
\end{center}

\vspace{\fill}

\begin{abstract}
We report a measurement of the lifetime of the $\Omegac$ baryon
using proton-proton collision data at center-of-mass energies of 7 and 8\tev, corresponding to an 
integrated luminosity of 3.0\invfb collected by the 
LHCb experiment. The sample
consists of about 1000 $\Omegab\to\Omegac\mun\neumb X$ signal decays, where the $\Omegac$ baryon is detected in the
$p\Km\Km\pip$ final state and $X$ represents possible additional undetected particles in the decay.
The $\Omegac$ lifetime is measured to be $\tau_{\Omegac} = 268\pm24\pm10\pm2$~{\rm fs}, where the uncertainties 
are statistical, systematic, and from the uncertainty in the $\Dp$ lifetime, respectively.
This value is nearly four times larger than, and inconsistent with, the current world-average value.
\end{abstract}

\vspace*{2.0cm}

\begin{center}
  Published in Phys. Rev. Lett. 121, 092003 (2018) 
\end{center}

\vspace{\fill}

{\footnotesize 
\centerline{\copyright~\papercopyright, \href{\paperlicenceurl}{\paperlicence} license.}}

\vspace*{2mm}

\end{titlepage}


\newpage
\setcounter{page}{2}
\mbox{~}
%
%
%
%

\cleardoublepage


\renewcommand{\thefootnote}{\arabic{footnote}}
\setcounter{footnote}{0}



\pagestyle{plain} 
\setcounter{page}{1}
\pagenumbering{arabic}


%

Measurements of the lifetimes of hadrons containing heavy ($b$ or $c$) quarks play an important role in testing theoretical approaches
that are used to predict Standard Model parameters. The validation of such tools is important, as they
can then be used to search for deviations from Standard Model expectations in other processes. One of the most predictive tools in quark flavor physics
is the heavy quark expansion (HQE)~\cite{Khoze:1983yp,Bigi:1991ir,Bigi:1992su,Blok:1992hw,Blok:1992he,Neubert:1997gu,Uraltsev:1998bk,Bigi:1995jr}, which
describes the decay widths of hadrons containing heavy quarks, $Q$, through an expansion in powers of 
$1/m_Q$, where $m_Q$ is the heavy quark mass. While predictions for absolute lifetimes carry relatively large uncertainties, ratios of lifetimes have smaller theoretical uncertainties~\cite{Lenz:2014jha}.
Higher-order terms in the HQE are related to non-perturbative corrections, and to effects due to the presence of the other 
light quark(s) (spectator) in the heavy hadron. For beauty hadrons with a single heavy quark, these corrections are
typically at the few percent level or less, due to the large mass of the $b$ quark~\cite{Lenz:2014jha}. For charm hadrons, since $m_c$ is significantly smaller than $m_b$, these higher-order corrections can be sizable. Therefore
measurements of charm-hadron lifetimes provide a sensitive probe of their contributions~\cite{Cheng:2015iom,Bianco:2003vb,Blok:1991st,Kirk:2017juj,Lenz:2013aua}.

While charm-meson lifetimes have been measured precisely and provide useful information on these
higher-order terms, the knowledge of charm-baryon lifetimes is much less accurate.  The lifetimes of the $\Dz$, $\Dp$ and $\Dsp$ mesons are 
known to about 1\% precision, whereas the corresponding uncertainties for the $\Lc$, $\Xicp$, $\Xicz$ and $\Omegac$ baryons are
3\%, 6\%, 10\% and 17\%, respectively~\cite{PDG2018}. Improved measurements of the charm-baryon lifetimes provide complementary information to what can be gleaned from
charm mesons. For example, contributions from $W$-exchange and constructive Pauli interference effects are present in charm-baryon decays, but are small or absent in charm-meson decays~\cite{Bianco:2003vb}. Moreover, for charm baryons,
the spectator system may have spin~0 ($\Lc$, $\Xicp$, $\Xicz$) or 
spin~1 ($\Omegac$), whereas for charm mesons,
the light quark spin is always equal to 1/2. 

It has been argued that the expected lifetime hierarchy, 
due to the higher order contributions discussed above, should be~\cite{Cheng:2015iom,Bianco:2003vb,Cheng:1997xba,Bellini:1996ra,Blok:1991st,Shifman:1984wx}
\begin{align}
\tau_{\Xicp} > \tau_{\Lc} > \tau_{\Xicz} > \tau_{\Omegac}.
\end{align}
\noindent The quark content of the $\Omegac$ baryon is $css$, and the qualitative argument that the $\Omegac$ lifetime should be the shortest is predicated on large constructive interference between the $s$ quark in the $c\to sW^+$ transition in the $\Omegac$ decay and the 
spectator $s$ quarks in the final state. However, it is also conceivable that the $\Omegac$ lifetime could be the largest, depending on the treatment of higher-order terms in the HQE expansion~\cite{Blok:1991st}.

Current measurements~\cite{PDG2018} are consistent with this hierarchy. The least well measured
lifetime is that of the $\Omegac$ baryon, with a value of $\tau_{\Omegac}=69\pm12$~fs, obtained by
fixed-target experiments using a small number of signal decays~\cite{Link:2003nq,Adamovich:1995pf,Frabetti:1995bi}.

In this Letter reports a new measurement of the $\Omegac$ baryon lifetime using a sample of 
semileptonic (SL) $\Omegab\to\Omegac\mun\neumb X$ decays, where the $\Omegac$ baryons are detected
in the $p\Km\Km\pip$ final state and $X$ represents any additional undetected particles. Semileptonic $b$-meson decays were used previously by LHCb to make precise measurements of the $\Ds$ and $\Bs$ lifetimes~\cite{LHCb-PAPER-2017-004}.
Throughout the text, charge-conjugate processes are implicitly included.

To reduce the uncertainties associated with systematic effects, the lifetime ratio
\begin{align}
r_{\Omegac}\equiv\frac{\tau_{\Omegac}}{\tau_{\Dp}}
\label{eq:rtau}
\end{align}
\noindent is measured, where the $\Dp$ meson is detected in $B\to\Dp\mun\neumb X$ decays, with \mbox{$\Dp\to\Km\pip\pip$}. 
In the following, the symbols $H_b$ and $H_c$ are used to refer to the $b$ or $c$-hadron in either of the two 
modes indicated above.

The measurement uses proton-proton ($pp$) collision data samples, collected by the LHCb experiment,
corresponding to an integrated luminosity of 3.0\invfb, of which 1.0\invfb was recorded at 
a center-of-mass energy of 7\tev and 2.0\invfb at 8\tev.  
The \lhcb detector~\cite{Alves:2008zz,LHCb-DP-2014-002} is a single-arm forward
spectrometer covering the \mbox{pseudorapidity} range $2<\eta <5$,
designed for the study of particles containing \bquark or \cquark
quarks. The tracking system provides a measurement of momentum, \ptot, of charged particles with
a relative uncertainty that varies from 0.5\% at low momentum to 1.0\% at 200\gevc.
The minimum distance of a track to a primary vertex (PV), the impact parameter (IP), 
is measured with a resolution of $(15+29/\pt)\mum$,
where \pt is the component of the momentum transverse to the beam, in\,\gevc.
Charged hadrons are identified using information
from two ring-imaging Cherenkov (RICH) detectors~\cite{LHCb-DP-2012-003}.
Muons are identified by a
system composed of alternating layers of iron and multiwire
proportional chambers~\cite{LHCb-DP-2013-001}. The online event selection is performed by a trigger~\cite{LHCb-DP-2012-004}, 
which consists of a hardware stage, based on information from the calorimeter and muon
systems, followed by a software stage, which applies a full event reconstruction.

Simulation is required to model the effects of the detector acceptance and the imposed 
selection requirements. Proton-proton collisions are simulated using
\pythia~\cite{Sjostrand:2006za,*Sjostrand:2007gs} with a specific \lhcb
configuration~\cite{LHCb-PROC-2010-056}.  Decays of hadronic particles
are described by \evtgen~\cite{Lange:2001uf}, in which final-state
radiation is generated using \photos~\cite{Golonka:2005pn}. The
interaction of the generated particles with the detector and its
response are implemented using the \geant toolkit~\cite{Allison:2006ve, *Agostinelli:2002hh} as described in
Ref.~\cite{LHCb-PROC-2011-006}. 

Signal $\Omegab$ candidates are formed by combining an $\Omegac\to p\Km\Km\pip$ candidate with a $\mun$ candidate. Each final-state particle in the decay is required to be detached from all PVs in the event, and is associated to the one with the smallest $\chisqip$. Here, $\chisqip$ is defined as the difference in $\chisq$ of the particle's associated PV reconstructed with and without the considered track. The muon is required to have $\pt>1\gevc$, $p>6\gevc$ and
have particle identification (PID) information consistent with being a muon.
The $\Omegac$ candidate's decay products must have PID information consistent with their assumed particle hypotheses, and have
$\pt>0.25\gevc$ and $p>2\gevc$, except for the proton, which is required to have $p>8\gevc$. 
To remove the contribution from promptly produced $\Omegac$
baryons, each $\Omegac$ candidate's reconstructed trajectory must not point back to any PV in the event.
Only $\Omegac$ candidates that have an invariant mass within 60\mevcc of the known $\Omegac$ mass are retained. 

The $\Omegac\mun$ combinations are required to form a good quality vertex and satisfy the invariant mass requirement, $m(\Omegac\mun)<8.0$\gevcc. 
Random combinations of $\Omegac$ and $\mun$ are suppressed by requiring the fitted $z$ coordinates of the $\Omegac$ and $\Omegab$ decay vertices to satisfy $z(\Omegac)-z(\Omegac\mun) > -0.05$~mm, where the $z$ axis is parallel to the beam direction. 

To ensure precise modeling of the decay-time acceptance from simulation, the candidates must satisfy a well-defined set of hardware and software trigger requirements. At the hardware level, candidates are required to
pass the single-muon trigger, and, at the software level, to pass specific triggers designed
to select multi-body final states containing a muon~\cite{LHCb-DP-2012-004}.

To improve the signal-to-background ratio in the $\Omegac\mun$ sample, a boosted decision tree (BDT) 
discriminant~\cite{Breiman, AdaBoost} is built from 18 variables, which include the $\chisq$ for the $\Omegab$ and $\Omegac$ 
decay-vertex fits, and $\chisqip$, $p$, $\pt$, and a PID response variable for each final-state hadron. The BDT is trained using simulated $\Omegab\to\Omegac\mun\neumb X$
decays for the signal, while background is taken from the $\Omegac$ mass sidebands, $30<|m(p\Km\Km\pip)-m_{\Omegac}|<50\mevcc$,
where $m_{\Omegac}$ is the known $\Omegac$ mass~\cite{PDG2018}.
The requirement on the BDT response is determined by optimizing the figure of merit $S/\sqrt{S+B}$,
where $S$ and $B$ are the expected signal and background yields within a $\pm$15\mevcc mass region centered on the mass peak, respectively. The optimal 
BDT requirement provides a signal (background) efficiency of 78\% (16\%).

The $\Dp\mun$ candidates, used for normalization, are formed by combining ${\Dp\to\Km\pip\pip}$ and $\mun$ candidates. The selections are identical to those
discussed above, except the mass window is centered on the known $\Dp$ mass and the BDT requirement is eliminated. Only 10\% of the $\Dp\mun$ data, selected at random, are used in the analysis,
since the full sample is much larger than needed for this measurement. 

\begin{figure}[tb]
\centering
\includegraphics[width=0.48\textwidth]{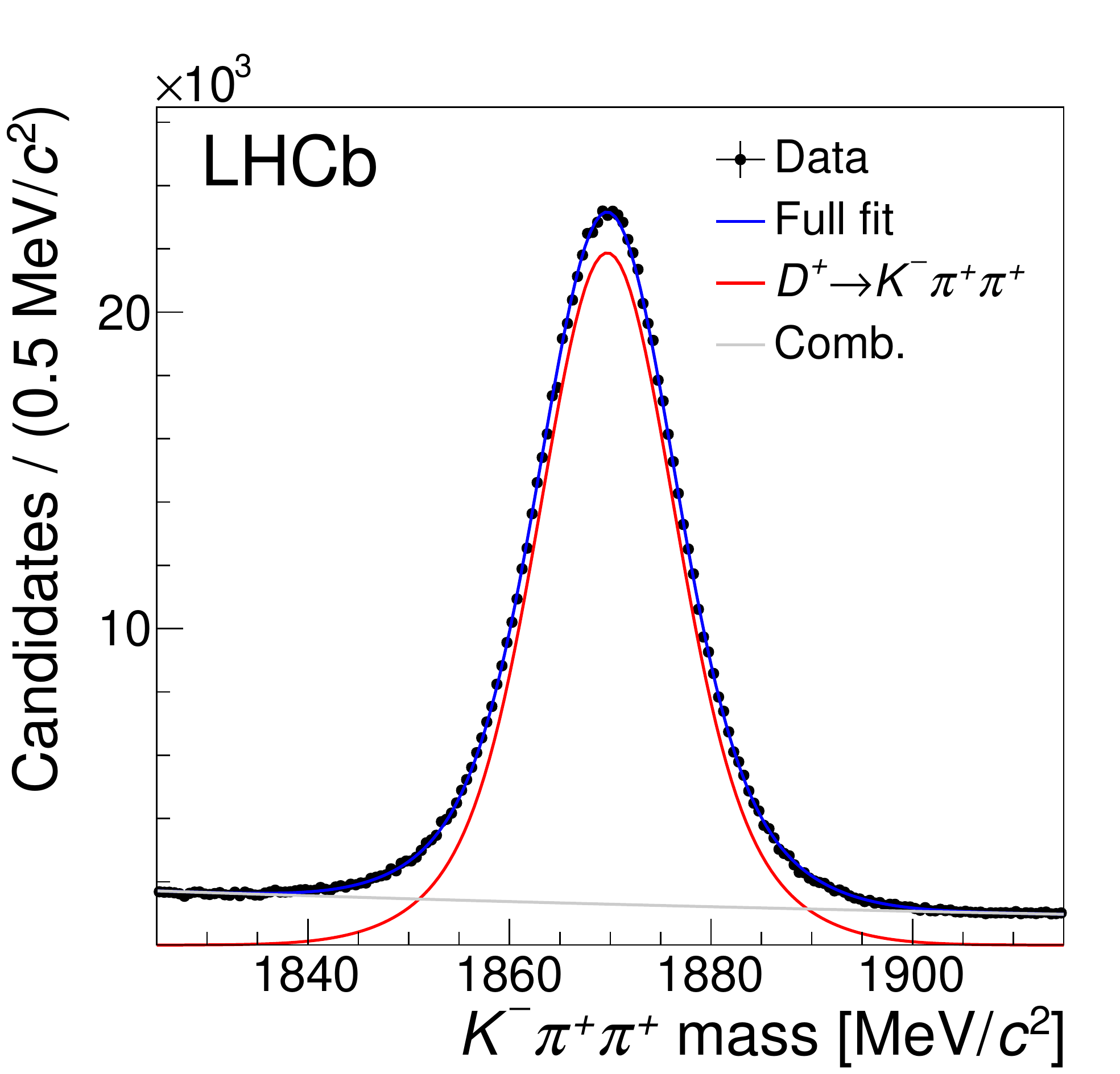}
\includegraphics[width=0.48\textwidth]{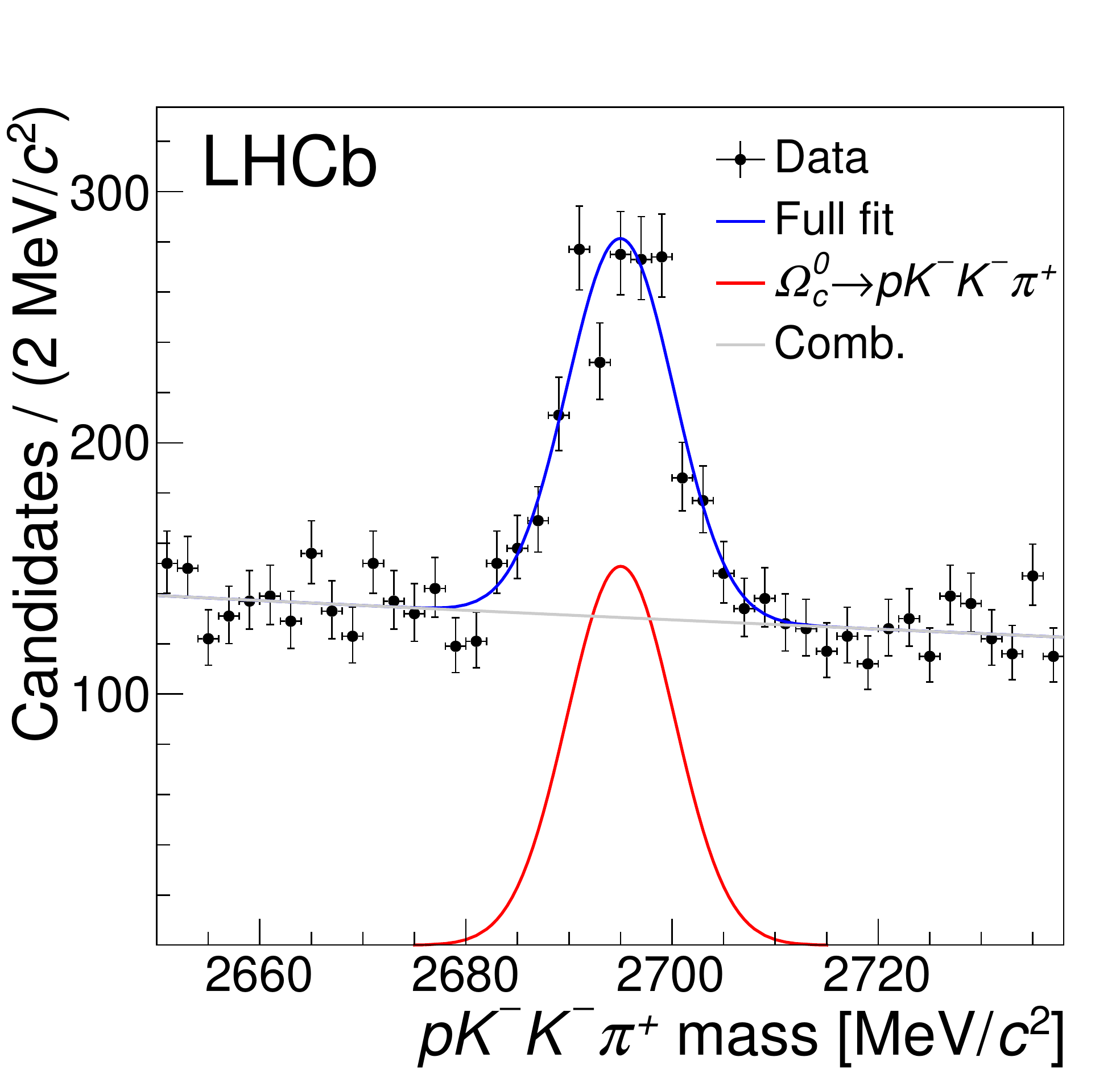}
\caption{\small{Invariant-mass distributions for (left) $\Dp$ candidates in $B\to\Dp\mun\neumb X$ decays
and (right) $\Omegac$ candidates in $\Omegab\to\Omegac\mun\neumb X$ decays.
The results of the fits, as described in the text, are overlaid.}}
\label{fig:MassPlots}
\end{figure}

The invariant-mass distributions for the selected $\Omegac$ and $\Dp$ candidates in the two $H_c\mun$ final states are shown in 
Fig.~\ref{fig:MassPlots}. Both distributions are fitted using the sum of a signal component, defined as the sum of two
Gaussian functions with a common mean, and an exponential shape to represent the combinatorial background.
From a binned maximum-likelihood fit, the fitted $\Omegac\mun$ and $\Dp\mun$ yields are $978\pm60$ and $(809\pm1)\times10^{3}$,
respectively. The number of $\Omegac$ signal decays
is at least an order of magnitude larger than any previous sample used for an $\Omegac$ lifetime measurement.

The decay time of each $H_c$ candidate is determined from the positions of the $H_b$ and $H_c$ decay vertices, and the measured 
$H_c$ momentum. The background-subtracted decay-time spectra are obtained using the \sPlot technique~\cite{Pivk:2004ty}, where the 
measured $H_c$ mass is used as the discriminating variable. 
The uncertainties in the bin-by-bin signal yields reflect both the finite signal yield and the statistical uncertainty due to the background subtraction.

Potential backgrounds from (i) random $H_c\mun$ combinations, 
(ii) $H_b\to H_c\taum\neutb$, \mbox{$\taum\to\mun\neut\neumb$} decays, and (iii) $H_b\to H_c\Dbar,~\Dbar\to\mun X$,
where $\Dbar$ represents a $\Dsm$, $\Dm$ or $\Dzb$ meson, could lead to a bias on the lifetime, since the
muon is not produced directly at the $H_b$ decay vertex.
These backgrounds have been investigated and constitute a small fraction of the observed signal, 
about 3\% in total, and have decay-time spectra that are similar to the true $H_c\mun\neumb$ final state due to the 
$\chisq$ requirements on the $H_b$ vertex fit. Moreover, these backgrounds affect the
signal and the normalization mode similarly, thus leading to at least a partial cancellation of any bias.
Contamination in the $\Omegab\to\Omegac\mun\neumb X$ sample from misidentified four-body $\Dz$ final states in 
$B\to\Dz\mun\neumb X$ decays has been investigated, and none are found to peak in the $\Omegac$ signal region.

\begin{figure}[tb]
\centering
\includegraphics[width=0.48\textwidth]{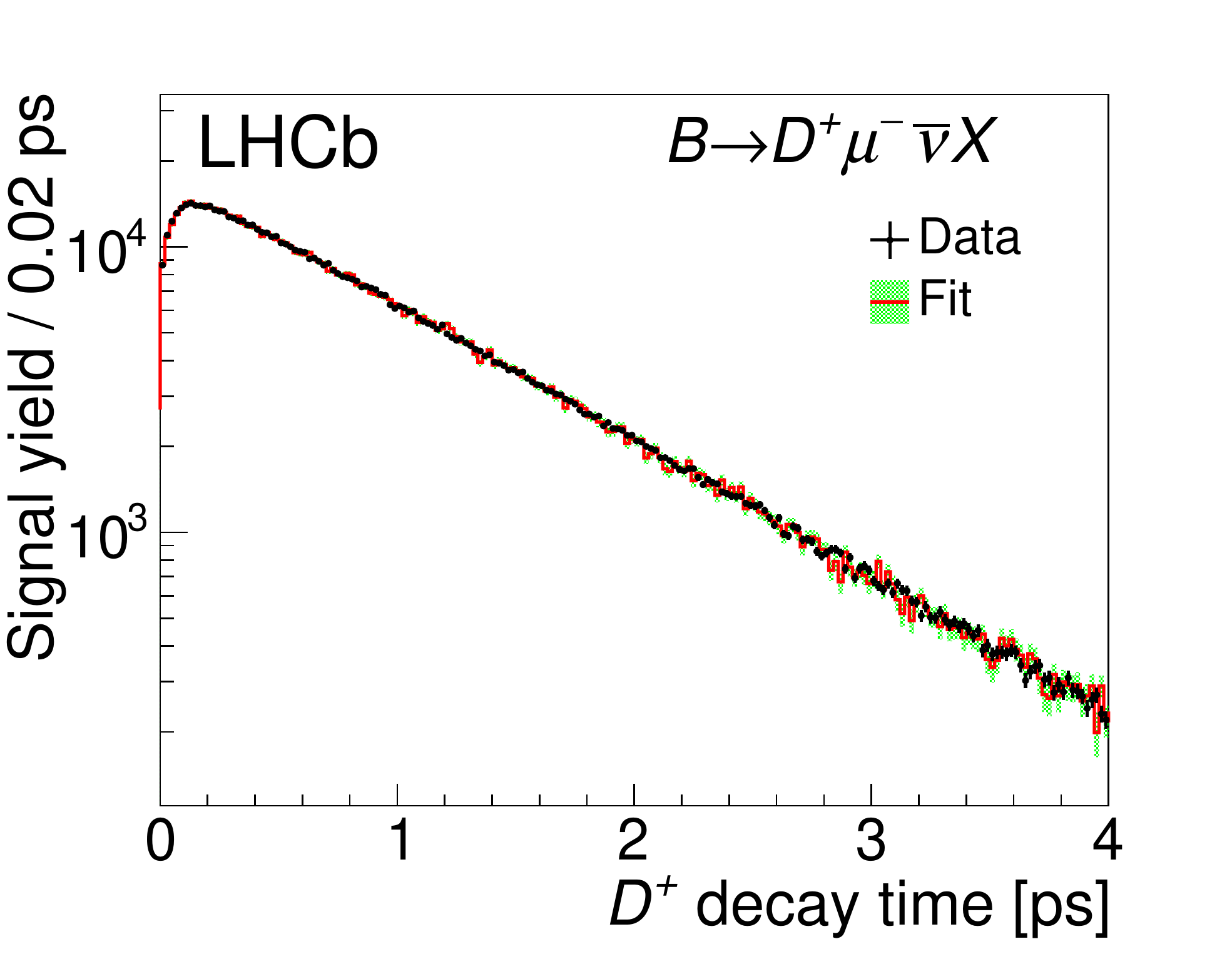}
\includegraphics[width=0.48\textwidth]{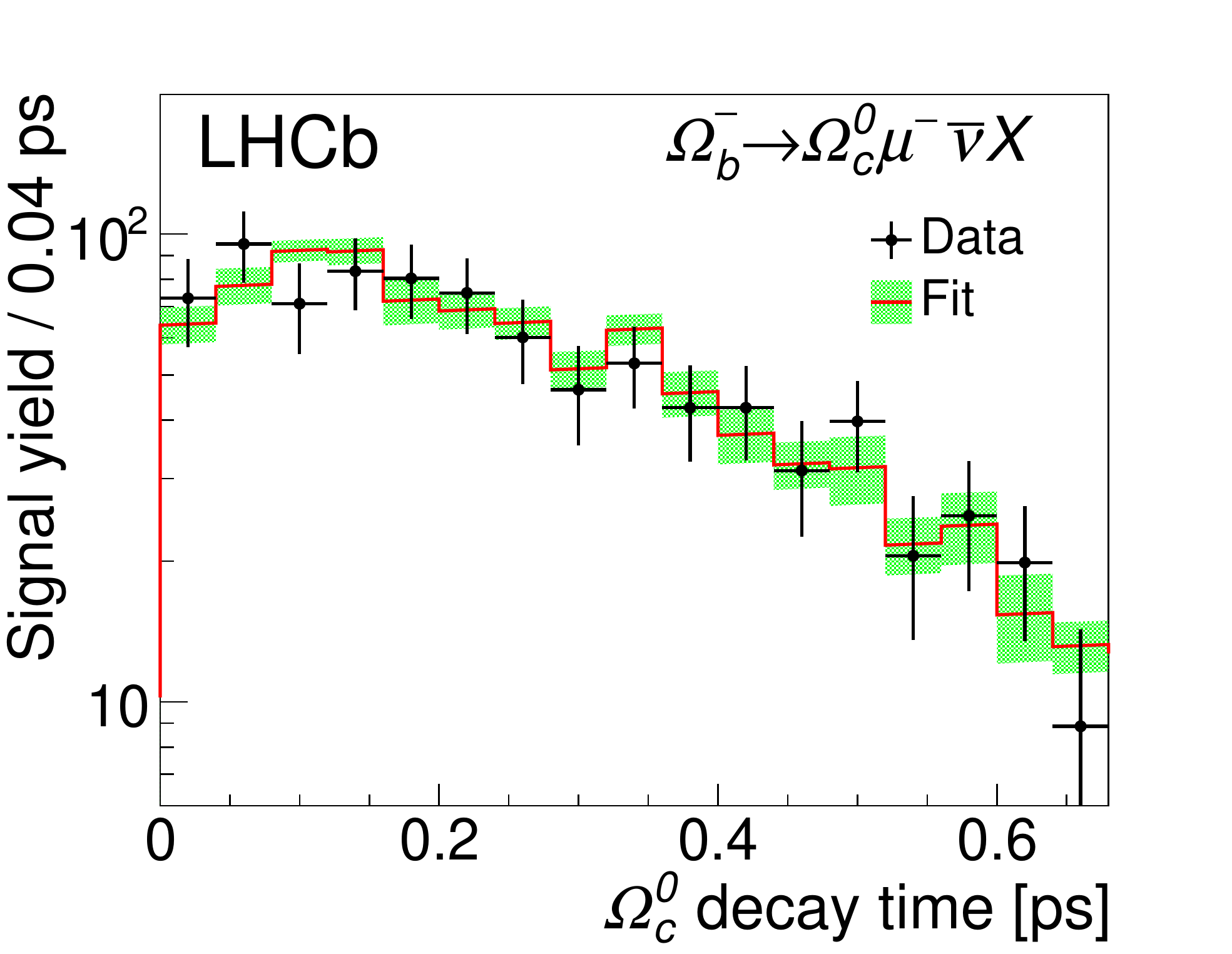}
\caption{\small{Decay-time spectra for (left) $\Dp$ signal in $B\to\Dp\mun X$ events and (right) $\Omegac$ signal in
$\Omegab\to\Omegac\mun X$ events. Overlaid are the fit results, as described in the text, along with the uncertainties due to finite simulated sample sizes.}}
\label{fig:XcDecaytimeRatioFits}
\end{figure}

The decay-time spectra for the $\Omegac$ and $\Dp$ signals are shown in Fig.~\ref{fig:XcDecaytimeRatioFits}, along 
with the results of the fits described below. The decrease in the signal yield as the decay time approaches zero is mainly due to the
effects of the $H_c$ decay-time resolution, which is in the range of $85-100$~fs, and the $z(H_c)-z(H_c\mun) > -0.05$~mm requirement.

The decay-time signal model, $S(t_{\rm rec})$, takes the form
\begin{align}
S(t_{\rm rec}) = f(t_{\rm rec}) g(t_{\rm rec}) \beta(t_{\rm rec}) . 
\end{align}
Here, $f(t_{\rm rec})$ is a signal template of reconstructed decay times, obtained from the full LHCb simulation,
after all selections have been applied as in the data.
The signal template is multiplied by $g(t_{\rm rec})=\exp(-t_{\rm rec}/\tau_{\rm fit}^{H_c}) / \exp(-t_{\rm rec}/\tau_{\rm sim}^{H_c})$,
where $\tau_{\rm sim}^{\Dp}=1040$~fs and $\tau_{\rm sim}^{\Omegac}=250$~fs are the lifetimes used in the simulation, 
and $\tau_{\rm fit}^{H_c}$ is the signal lifetime to be fitted.
The function $\beta(t_{\rm rec})$ is a correction that accounts for a small difference in the efficiency
between data and simulation for reconstructing tracks in the vertex detector that originate far from the beamline~\cite{LHCb-PAPER-2013-065}. 

Given the precise knowledge of the $\Dp$ meson lifetime ($1040\pm7$~fs)~\cite{PDG2018}, the $\Dp\mun$ sample is used to
calibrate $\beta(t_{\rm rec})$ and validate the fit. The signal template is obtained from simulated $B\to\Dp\mun\neumb X$ decays,
where contributions from $B\to\Dp\taum\neutb X$ decays are included. The function $\beta(t_{\rm rec})$ is obtained by 
taking the ratio between the $\Dp$ decay-time spectrum in data (obtained via the \sPlot technique) and that obtained from 
simulation. The ratio shows a linear dependence, and a fit to the function $\beta(t_{\rm rec}) = 1+\beta_0 t_{\rm rec}$ yields
$\beta_0 = (-0.89\pm0.32)\times10^{-2}$~ps$^{-1}$. If the $\beta(t_{\rm rec})$ function is excluded from the fit, 
$\tau^{\Dp}_{\rm fit}$ is $10$~fs below the world average.
The result of the binned $\chi^2$ fit after this correction is applied is shown in
Fig.~\ref{fig:XcDecaytimeRatioFits}\,(left), where the fitted lifetime is found to be $\tau_{\rm fit}^{\Dp}=1042.0\pm1.7\stat$~fs.

The $\Omegac$ lifetime is determined from a simultaneous fit to the $\Omegac$ and $\Dp$ decay-time spectra, for which the free parameters in the fit are $r_{\Omegac}$ (see Eq.~\ref{eq:rtau}) and $\tau_{\rm fit}^{\Dp}$. By fitting for the
ratio $r_{\Omegac}$, correlated systematic uncertainties partially cancel.
In the $\Omegac$ decay-time fit, $\beta_0$ is scaled by 4/3 since the effect is expected to scale with the number of charged final 
state particles in the $H_c$ decay~\cite{LHCb-PAPER-2013-065}.
The simulation includes contributions from $\Omegac\taum\neutb X$ final states.
The results of the fit to the $\Omegac$ decay-time distribution are shown in Fig.~\ref{fig:XcDecaytimeRatioFits}\,(right), where the value
$r_{\Omegac}=0.258\pm0.023\stat$ is obtained. Multiplying this value by $\tau^{\Dp}=1040$~fs~\cite{PDG2018}, the
$\Omegac$ lifetime is measured to be $268\pm24$~fs. This is about four times larger than, and incompatible with, the current
world average value of $69\pm12$~fs~\cite{PDG2018}.

Several cross-checks have been performed to ensure the robustness of this result. To confirm that the signal events
are from SL $\Omegab$ decays, a number of distributions, such as the $\Omegac\mun$ mass spectrum, $\pt$ and decay time
have been compared between data (using \sPlot) and the $\Omegab\to\Omegac\mun\neumb X$ simulation. In all cases,
good agreement is found. The lifetime measurement has also been performed using a simple subtraction of the $\Omegac$ mass sidebands,
and we find good agreement with the value obtained by the \sPlot technique. The $\Omegac$ decay-time distribution obtained from an independent 
and comparably sized data sample of semileptonic decays collected at 13\tev center-of-mass energy
has been examined, and the distribution is consistent with the one observed here. 
The procedure has also been checked using a sample of about 88,000 $\Bm\to\Dz(\to\Kp\Km\pip\pim)\mun X$
decays to measure the $\Dz$ meson lifetime. The obtained lifetime is consistent with the expected value within about one standard deviation.
The analysis has also been carried out with either tighter PID or tighter BDT requirements, and the fitted $\Omegac$ lifetime in each case is consistent with
the value from the default fit. The analysis has also been checked with the $\Lc$ baryon, and the lifetime is 
consistent with expectations. 

A number of sources of systematic uncertainty on the measured ratio $r_{\Omegac}$ have been investigated, and are summarized in Table~\ref{tab:CharmSyst}.
The decay time acceptance correction, $\beta(t_{\rm rec})$, leads to an uncertainty of 0.5\% on $r_{\Omegac}$, which includes a contribution from the finite sample sizes and the choice of fit function. 

Studies of the $\Dp$ calibration mode show a small dependence of the $\beta_0$ parameter on the $\pt$ and $\eta$ of the $H_b$ hadron. In the case that the $\pt$ and $\eta$ spectra in data and simulation differ, it could cause
a shift in the average $\beta_0$. 
The uncertainty on $r_{\Omegac}$ is obtained by taking into account the variation of $\beta_0$ in different 
$\pt$ and $\eta$ ranges, and the extent to which the $\pt$ and $\eta$ spectra may differ between data and simulation.

The world-average value of the $\Omegab$ lifetime is $1.64^{+0.18}_{-0.17}$~ps~\cite{PDG2018}, whereas the simulation uses 1.60~ps. 
To assess the potential impact on the $\Omegac$ lifetime, we weight $f(t_{\rm rec})$ to replicate an $\Omegab$ lifetime of either 
1.50~ps or 1.70~ps. The changes in $r_{\Omegac}$ are assigned as a systematic uncertainty.

The decay-time resolution has been checked by comparing the $\Dz$ decay-time spectra in
$\Bm\to\Dz\pim$ decays, where no explicit requirement on the flight distance of the $\Dz$ is applied. 
Negative decay times are entirely due to the decay-time resolution, and simulation is found to agree well
with data. To assess a potential impact of a small difference
in decay-time resolution between simulation and data, new $\Omegac$ and $\Dp$ signal templates are formed where
the reconstructed decay time is smeared by an additional 15\%, beyond what is produced by the full simulation.
The fit is redone, and the difference in $r_{\Omegac}$ from the nominal value is assigned as a systematic uncertainty.

The method for background subtraction uses the \sPlot technique, which has some dependence on the choice of
signal and background functions. 
To assess a potential systematic effect, the decay-time 
spectra are obtained using a sideband subtraction of the $H_c$ mass spectra for both the signal and the 
normalization modes. 
The sideband-subtracted decay-time spectra are then fitted 
using the decay-time fit described above. The difference between this result and the nominal one is assigned as a systematic 
uncertainty. 

The decay-time spectra in both $\Omegac\mun$ and $\Dp\mun$ samples, have small contributions from random combinations of
$H_c$ and $\mun$ candidates [($0.8\pm0.2)$\% of the signal], as well as physics backgrounds where the $\mun$ comes from either a $\taum$ [($1.8\pm0.3$\%)] or a SL $D$ decay [($0.5\pm0.2)$\%]. 
From simulation and data control samples, we find that the effective lifetimes of these backgrounds
are within 10\% of the true signal lifetime; this is due to
the requirement that the muon candidate must form a good vertex with the $H_c$ candidate. The impact on the $\Omegac$ lifetime 
is evaluated using pseudoexperiments, where mixtures of these backgrounds (with different decay-time spectra)
and signal decays are formed and fitted assuming a single lifetime for the sample. The difference in the mean value of $r_{\Omegac}$ between the nominal fit, and that with the backgrounds added is assigned as the systematic uncertainty. 

The systematic uncertainty due to the finite size of the simulated samples is assessed by repeating the fit to the data many times, where in each fit the
simulated-template bin contents are fluctuated within their uncertainties. The standard deviation of the distribution of the
fitted $r_{\Omegac}$ values is assigned as a systematic uncertainty.  

\begin{table*}[tb]
\begin{center}
\caption{\small{Summary of systematic uncertainties on the lifetime ratio, $r_{\Omegac}$, in units of $10^{-4}$.}}
\begin{tabular}{lc}
\hline\hline
Source                         &  $r_{\Omegac}$ \rule[-1.2ex]{0pt}{0 pt} ($10^{-4}$)\\

\hline  
Decay-time acceptance          &    ~13         \\
$\Omegab$ prod. spectrum       &    ~~3     \\
$\Omegab$ lifetime             &    ~~4    \\
Decay-time resolution          &    ~~3         \\
Background subtraction         &    ~18       \\
$H_c(\taum,D)$, random $\mun$ &    ~~8          \\
Simulated sample size         &    ~98       \\
\hline
Total systematic              &    101      \\
\hline
Statistical uncertainty       &   230      \\
\hline\hline
\end{tabular}
\label{tab:CharmSyst}
\end{center}
\end{table*}

In summary, we use $pp$ collision data samples at 7\tev and 8\tev center of mass energies,
corresponding to 3.0\invfb of integrated luminosity,
to measure the lifetime of the $\Omegac$ baryon. The measured ratio of lifetimes and absolute $\Omegac$ lifetime are
\begin{align*}
  \frac{\tau_{\Omegac}}{\tau_{\Dp}} &= 0.258\pm0.023\pm0.010 \\
  \tau_{\Omegac} &= 268\pm24\pm10\pm2~{\rm fs},
\end{align*}
\noindent where the first uncertainty is statistical, the second is systematic, and the third is
due to the uncertainty in the $\Dp$ lifetime~\cite{PDG2018}. The measured $\Omegac$ lifetime is about four times larger than, and inconsistent
with, the world average value of $69\pm12$~fs~\cite{PDG2018}. 

With this measurement, the lifetime hierarchy places the $\Omegac$ baryon as having the second largest lifetime 
after the $\Xicp$ baryon,
\begin{align*}
\tau_{\Xicp} > \tau_{\Omegac} > \tau_{\Lc} > \tau_{\Xicz}.
\end{align*}
The result presented here may suggest that the constructive interference between the $s$ quark in the $c\to sW^+$ transition in the $\Omegac$ decay and the 
spectator $s$ quarks in the final state is smaller than expected, that the spin of the $ss$ system plays a larger role, or that additional or higher order contributions in the heavy quark expansion need to be considered.


\section*{Acknowledgements}
%
%
\noindent We express our gratitude to our colleagues in the CERN
accelerator departments for the excellent performance of the LHC. We
thank the technical and administrative staff at the LHCb
institutes.
We acknowledge support from CERN and from the national agencies:
CAPES, CNPq, FAPERJ and FINEP (Brazil);
MOST and NSFC (China);
CNRS/IN2P3 (France);
BMBF, DFG and MPG (Germany);
INFN (Italy);
NWO (Netherlands);
MNiSW and NCN (Poland);
MEN/IFA (Romania);
MinES and FASO (Russia);
MinECo (Spain);
SNSF and SER (Switzerland);
NASU (Ukraine);
STFC (United Kingdom);
NSF (USA).
We acknowledge the computing resources that are provided by CERN, IN2P3
(France), KIT and DESY (Germany), INFN (Italy), SURF (Netherlands),
PIC (Spain), GridPP (United Kingdom), RRCKI and Yandex
LLC (Russia), CSCS (Switzerland), IFIN-HH (Romania), CBPF (Brazil),
PL-GRID (Poland) and OSC (USA).
We are indebted to the communities behind the multiple open-source
software packages on which we depend.
Individual groups or members have received support from
AvH Foundation (Germany);
EPLANET, Marie Sk\l{}odowska-Curie Actions and ERC (European Union);
ANR, Labex P2IO and OCEVU, and R\'{e}gion Auvergne-Rh\^{o}ne-Alpes (France);
Key Research Program of Frontier Sciences of CAS, CAS PIFI, and the Thousand Talents Program (China);
RFBR, RSF and Yandex LLC (Russia);
GVA, XuntaGal and GENCAT (Spain);
Herchel Smith Fund, the Royal Society, the English-Speaking Union and the Leverhulme Trust (United Kingdom);
Laboratory Directed Research and Development program of LANL (USA).

\clearpage
\clearpage

\clearpage

\addcontentsline{toc}{section}{References}
\setboolean{inbibliography}{true}
\bibliographystyle{LHCb}
\bibliography{main,LHCb-PAPER,LHCb-CONF,LHCb-DP,LHCb-TDR,steve}

\ifx\mcitethebibliography\mciteundefinedmacro
\PackageError{LHCb.bst}{mciteplus.sty has not been loaded}
{This bibstyle requires the use of the mciteplus package.}\fi
\providecommand{\href}[2]{#2}
\begin{mcitethebibliography}{10}
\mciteSetBstSublistMode{n}
\mciteSetBstMaxWidthForm{subitem}{\alph{mcitesubitemcount})}
\mciteSetBstSublistLabelBeginEnd{\mcitemaxwidthsubitemform\space}
{\relax}{\relax}

\bibitem{Khoze:1983yp}
V.~A. Khoze and M.~A. Shifman,
  \ifthenelse{\boolean{articletitles}}{\emph{{Heavy quarks}},
  }{}\href{https://doi.org/10.1070/PU1983v026n05ABEH004398}{Sov.\ Phys.\ Usp.\
  \textbf{26} (1983) 387}\relax
\mciteBstWouldAddEndPuncttrue
\mciteSetBstMidEndSepPunct{\mcitedefaultmidpunct}
{\mcitedefaultendpunct}{\mcitedefaultseppunct}\relax
\EndOfBibitem
\bibitem{Bigi:1991ir}
I.~I.~Y. Bigi and N.~G. Uraltsev,
  \ifthenelse{\boolean{articletitles}}{\emph{{Gluonic enhancements in
  non-spectator beauty decays - an inclusive mirage though an exclusive
  possibility}}, }{}\href{https://doi.org/10.1016/0370-2693(92)90066-D}{Phys.\
  Lett.\  \textbf{B280} (1992) 271}\relax
\mciteBstWouldAddEndPuncttrue
\mciteSetBstMidEndSepPunct{\mcitedefaultmidpunct}
{\mcitedefaultendpunct}{\mcitedefaultseppunct}\relax
\EndOfBibitem
\bibitem{Bigi:1992su}
I.~I. Bigi, N.~G. Uraltsev, and A.~I. Vainshtein,
  \ifthenelse{\boolean{articletitles}}{\emph{{Nonperturbative corrections to
  inclusive beauty and charm decays. QCD versus phenomenological models}},
  }{}\href{https://doi.org/10.1016/0370-2693(92)90908-M}{Phys.\ Lett.\
  \textbf{B293} (1992) 430}, Erratum
  \href{https://doi.org/10.1016/0370-2693(92)91287-J}{ibid.\   \textbf{B297}
  (1992) 477}, \href{http://arxiv.org/abs/hep-ph/9207214}{{\normalfont\ttfamily
  arXiv:hep-ph/9207214}}\relax
\mciteBstWouldAddEndPuncttrue
\mciteSetBstMidEndSepPunct{\mcitedefaultmidpunct}
{\mcitedefaultendpunct}{\mcitedefaultseppunct}\relax
\EndOfBibitem
\bibitem{Blok:1992hw}
B.~Blok and M.~Shifman, \ifthenelse{\boolean{articletitles}}{\emph{{The rule of
  discarding 1/$N_c$ in inclusive weak decays (I)}},
  }{}\href{https://doi.org/10.1016/0550-3213(93)90504-I}{Nucl.\ Phys.\
  \textbf{B399} (1993) 441},
  \href{http://arxiv.org/abs/hep-ph/9207236}{{\normalfont\ttfamily
  arXiv:hep-ph/9207236}}\relax
\mciteBstWouldAddEndPuncttrue
\mciteSetBstMidEndSepPunct{\mcitedefaultmidpunct}
{\mcitedefaultendpunct}{\mcitedefaultseppunct}\relax
\EndOfBibitem
\bibitem{Blok:1992he}
B.~Blok and M.~Shifman, \ifthenelse{\boolean{articletitles}}{\emph{{The rule of
  discarding 1/$N_c$ in inclusive weak decays (II)}},
  }{}\href{https://doi.org/10.1016/0550-3213(93)90505-J}{Nucl.\ Phys.\
  \textbf{B399} (1993) 459},
  \href{http://arxiv.org/abs/hep-ph/9209289}{{\normalfont\ttfamily
  arXiv:hep-ph/9209289}}\relax
\mciteBstWouldAddEndPuncttrue
\mciteSetBstMidEndSepPunct{\mcitedefaultmidpunct}
{\mcitedefaultendpunct}{\mcitedefaultseppunct}\relax
\EndOfBibitem
\bibitem{Neubert:1997gu}
M.~Neubert, \ifthenelse{\boolean{articletitles}}{\emph{{B decays and the heavy
  quark expansion}}, }{}\href{https://doi.org/10.1142/9789812812667_0003}{Adv.\
  Ser.\ Direct.\ High Energy Phys.\  \textbf{15} (1998) 239},
  \href{http://arxiv.org/abs/hep-ph/9702375}{{\normalfont\ttfamily
  arXiv:hep-ph/9702375}}\relax
\mciteBstWouldAddEndPuncttrue
\mciteSetBstMidEndSepPunct{\mcitedefaultmidpunct}
{\mcitedefaultendpunct}{\mcitedefaultseppunct}\relax
\EndOfBibitem
\bibitem{Uraltsev:1998bk}
N.~Uraltsev, \ifthenelse{\boolean{articletitles}}{\emph{{Heavy quark expansion
  in beauty and its decays}},
  }{}\href{http://arxiv.org/abs/hep-ph/9804275}{{\normalfont\ttfamily
  arXiv:hep-ph/9804275}}, also published in proceedings, Heavy Flavour Physics:
  A Probe of Nature's Grand Design, Proc. Intern. School of Physics ``Enrico
  Fermi'', Course CXXXVII, Varenna, July 7-18 1997\relax
\mciteBstWouldAddEndPuncttrue
\mciteSetBstMidEndSepPunct{\mcitedefaultmidpunct}
{\mcitedefaultendpunct}{\mcitedefaultseppunct}\relax
\EndOfBibitem
\bibitem{Bigi:1995jr}
I.~I. Bigi, \ifthenelse{\boolean{articletitles}}{\emph{{The QCD perspective on
  lifetimes of heavy flavor hadrons}},
  }{}\href{http://arxiv.org/abs/hep-ph/9508408}{{\normalfont\ttfamily
  arXiv:hep-ph/9508408}}\relax
\mciteBstWouldAddEndPuncttrue
\mciteSetBstMidEndSepPunct{\mcitedefaultmidpunct}
{\mcitedefaultendpunct}{\mcitedefaultseppunct}\relax
\EndOfBibitem
\bibitem{Lenz:2014jha}
A.~Lenz, \ifthenelse{\boolean{articletitles}}{\emph{{Lifetimes and heavy quark
  expansion}}, }{}\href{https://doi.org/10.1142/S0217751X15430058}{Int.\ J.\
  Mod.\ Phys.\  \textbf{A30} (2015) 1543005},
  \href{http://arxiv.org/abs/1405.3601}{{\normalfont\ttfamily
  arXiv:1405.3601}}\relax
\mciteBstWouldAddEndPuncttrue
\mciteSetBstMidEndSepPunct{\mcitedefaultmidpunct}
{\mcitedefaultendpunct}{\mcitedefaultseppunct}\relax
\EndOfBibitem
\bibitem{Cheng:2015iom}
H.-Y. Cheng, \ifthenelse{\boolean{articletitles}}{\emph{{Charmed baryons circa
  2015}}, }{}\href{https://doi.org/10.1007/s11467-015-0483-z}{Front.\ Phys.\
  (Beijing) \textbf{10} (2015) 101406},
  \href{http://arxiv.org/abs/1508.07233}{{\normalfont\ttfamily
  arXiv:1508.07233}}\relax
\mciteBstWouldAddEndPuncttrue
\mciteSetBstMidEndSepPunct{\mcitedefaultmidpunct}
{\mcitedefaultendpunct}{\mcitedefaultseppunct}\relax
\EndOfBibitem
\bibitem{Bianco:2003vb}
S.~Bianco, F.~L. Fabbri, D.~Benson, and I.~Bigi,
  \ifthenelse{\boolean{articletitles}}{\emph{{A cicerone for the physics of
  charm}}, }{}\href{https://doi.org/10.1393/ncr/i2003-10003-1}{Riv.\ Nuovo
  Cim.\  \textbf{26N7} (2003) 1},
  \href{http://arxiv.org/abs/hep-ex/0309021}{{\normalfont\ttfamily
  arXiv:hep-ex/0309021}}\relax
\mciteBstWouldAddEndPuncttrue
\mciteSetBstMidEndSepPunct{\mcitedefaultmidpunct}
{\mcitedefaultendpunct}{\mcitedefaultseppunct}\relax
\EndOfBibitem
\bibitem{Blok:1991st}
B.~Blok and M.~A. Shifman,
  \ifthenelse{\boolean{articletitles}}{\emph{{Lifetimes of charmed hadrons
  revisited. Facts and fancy}}, }{} in {\em {Tau charm factory. Proceedings,
  3rd Workshop, Marbella, Spain, June 1-6, 1993}}, 1991.
\newblock \href{http://arxiv.org/abs/hep-ph/9311331}{{\normalfont\ttfamily
  arXiv:hep-ph/9311331}}\relax
\mciteBstWouldAddEndPuncttrue
\mciteSetBstMidEndSepPunct{\mcitedefaultmidpunct}
{\mcitedefaultendpunct}{\mcitedefaultseppunct}\relax
\EndOfBibitem
\bibitem{Kirk:2017juj}
M.~Kirk, A.~Lenz, and T.~Rauh,
  \ifthenelse{\boolean{articletitles}}{\emph{{Dimension-six matrix elements for
  meson mixing and lifetimes from sum rules}},
  }{}\href{https://doi.org/10.1007/JHEP12(2017)068}{JHEP \textbf{12} (2017)
  068}, \href{http://arxiv.org/abs/1711.02100}{{\normalfont\ttfamily
  arXiv:1711.02100}}\relax
\mciteBstWouldAddEndPuncttrue
\mciteSetBstMidEndSepPunct{\mcitedefaultmidpunct}
{\mcitedefaultendpunct}{\mcitedefaultseppunct}\relax
\EndOfBibitem
\bibitem{Lenz:2013aua}
A.~Lenz and T.~Rauh, \ifthenelse{\boolean{articletitles}}{\emph{{D-meson
  lifetimes within the heavy quark expansion}},
  }{}\href{https://doi.org/10.1103/PhysRevD.88.034004}{Phys.\ Rev.\
  \textbf{D88} (2013) 034004},
  \href{http://arxiv.org/abs/1305.3588}{{\normalfont\ttfamily
  arXiv:1305.3588}}\relax
\mciteBstWouldAddEndPuncttrue
\mciteSetBstMidEndSepPunct{\mcitedefaultmidpunct}
{\mcitedefaultendpunct}{\mcitedefaultseppunct}\relax
\EndOfBibitem
\bibitem{PDG2018}
Particle Data Group, M.~Tanabashi {\em et~al.},
  \ifthenelse{\boolean{articletitles}}{\emph{{\href{http://pdg.lbl.gov/}{Review
  of particle physics}}}, }{}Phys.\ Rev.\  \textbf{D98} (2018) 030001\relax
\mciteBstWouldAddEndPuncttrue
\mciteSetBstMidEndSepPunct{\mcitedefaultmidpunct}
{\mcitedefaultendpunct}{\mcitedefaultseppunct}\relax
\EndOfBibitem
\bibitem{Cheng:1997xba}
H.-Y. Cheng, \ifthenelse{\boolean{articletitles}}{\emph{{A phenomenological
  analysis of heavy hadron lifetimes}},
  }{}\href{https://doi.org/10.1103/PhysRevD.56.2783}{Phys.\ Rev.\  \textbf{D56}
  (1997) 2783},
  \href{http://arxiv.org/abs/hep-ph/9704260}{{\normalfont\ttfamily
  arXiv:hep-ph/9704260}}\relax
\mciteBstWouldAddEndPuncttrue
\mciteSetBstMidEndSepPunct{\mcitedefaultmidpunct}
{\mcitedefaultendpunct}{\mcitedefaultseppunct}\relax
\EndOfBibitem
\bibitem{Bellini:1996ra}
G.~Bellini, I.~I.~Y. Bigi, and P.~J. Dornan,
  \ifthenelse{\boolean{articletitles}}{\emph{{Lifetimes of charm and beauty
  hadrons}}, }{}\href{https://doi.org/10.1016/S0370-1573(97)00005-7}{Phys.\
  Rept.\  \textbf{289} (1997) 1}\relax
\mciteBstWouldAddEndPuncttrue
\mciteSetBstMidEndSepPunct{\mcitedefaultmidpunct}
{\mcitedefaultendpunct}{\mcitedefaultseppunct}\relax
\EndOfBibitem
\bibitem{Shifman:1984wx}
M.~A. Shifman and M.~B. Voloshin,
  \ifthenelse{\boolean{articletitles}}{\emph{{Preasymptotic effects in
  inclusive weak decays of charmed particles}}, }{}Sov.\ J.\ Nucl.\ Phys.\
  \textbf{41} (1985) 120, [Yad. Fiz.41,187(1985)]\relax
\mciteBstWouldAddEndPuncttrue
\mciteSetBstMidEndSepPunct{\mcitedefaultmidpunct}
{\mcitedefaultendpunct}{\mcitedefaultseppunct}\relax
\EndOfBibitem
\bibitem{Link:2003nq}
FOCUS collaboration, J.~M. Link {\em et~al.},
  \ifthenelse{\boolean{articletitles}}{\emph{{Measurement of the $\Omegac$
  lifetime}}, }{}\href{https://doi.org/10.1016/S0370-2693(03)00388-5}{Phys.\
  Lett.\  \textbf{B561} (2003) 41},
  \href{http://arxiv.org/abs/hep-ex/0302033}{{\normalfont\ttfamily
  arXiv:hep-ex/0302033}}\relax
\mciteBstWouldAddEndPuncttrue
\mciteSetBstMidEndSepPunct{\mcitedefaultmidpunct}
{\mcitedefaultendpunct}{\mcitedefaultseppunct}\relax
\EndOfBibitem
\bibitem{Adamovich:1995pf}
WA89 collaboration, M.~I. Adamovich {\em et~al.},
  \ifthenelse{\boolean{articletitles}}{\emph{{Measurement of the $\Omegac$
  lifetime}}, }{}\href{https://doi.org/10.1016/0370-2693(95)00979-U}{Phys.\
  Lett.\  \textbf{B358} (1995) 151},
  \href{http://arxiv.org/abs/hep-ex/9507004}{{\normalfont\ttfamily
  arXiv:hep-ex/9507004}}\relax
\mciteBstWouldAddEndPuncttrue
\mciteSetBstMidEndSepPunct{\mcitedefaultmidpunct}
{\mcitedefaultendpunct}{\mcitedefaultseppunct}\relax
\EndOfBibitem
\bibitem{Frabetti:1995bi}
E687 collaboration, P.~L. Frabetti {\em et~al.},
  \ifthenelse{\boolean{articletitles}}{\emph{{First measurement of the lifetime
  of the $\Omegac$}},
  }{}\href{https://doi.org/10.1016/0370-2693(95)00941-D}{Phys.\ Lett.\
  \textbf{B357} (1995) 678}\relax
\mciteBstWouldAddEndPuncttrue
\mciteSetBstMidEndSepPunct{\mcitedefaultmidpunct}
{\mcitedefaultendpunct}{\mcitedefaultseppunct}\relax
\EndOfBibitem
\bibitem{LHCb-PAPER-2017-004}
LHCb collaboration, R.~Aaij {\em et~al.},
  \ifthenelse{\boolean{articletitles}}{\emph{{Measurement of $\Bs$ and $\Dsm$
  meson lifetimes}},
  }{}\href{https://doi.org/10.1103/PhysRevLett.119.101801}{Phys.\ Rev.\ Lett.\
  \textbf{119} (2017) 101801},
  \href{http://arxiv.org/abs/1705.03475}{{\normalfont\ttfamily
  arXiv:1705.03475}}\relax
\mciteBstWouldAddEndPuncttrue
\mciteSetBstMidEndSepPunct{\mcitedefaultmidpunct}
{\mcitedefaultendpunct}{\mcitedefaultseppunct}\relax
\EndOfBibitem
\bibitem{Alves:2008zz}
LHCb collaboration, A.~A. Alves~Jr.\ {\em et~al.},
  \ifthenelse{\boolean{articletitles}}{\emph{{The \lhcb detector at the LHC}},
  }{}\href{https://doi.org/10.1088/1748-0221/3/08/S08005}{JINST \textbf{3}
  (2008) S08005}\relax
\mciteBstWouldAddEndPuncttrue
\mciteSetBstMidEndSepPunct{\mcitedefaultmidpunct}
{\mcitedefaultendpunct}{\mcitedefaultseppunct}\relax
\EndOfBibitem
\bibitem{LHCb-DP-2014-002}
LHCb collaboration, R.~Aaij {\em et~al.},
  \ifthenelse{\boolean{articletitles}}{\emph{{LHCb detector performance}},
  }{}\href{https://doi.org/10.1142/S0217751X15300227}{Int.\ J.\ Mod.\ Phys.\
  \textbf{A30} (2015) 1530022},
  \href{http://arxiv.org/abs/1412.6352}{{\normalfont\ttfamily
  arXiv:1412.6352}}\relax
\mciteBstWouldAddEndPuncttrue
\mciteSetBstMidEndSepPunct{\mcitedefaultmidpunct}
{\mcitedefaultendpunct}{\mcitedefaultseppunct}\relax
\EndOfBibitem
\bibitem{LHCb-DP-2012-003}
M.~Adinolfi {\em et~al.},
  \ifthenelse{\boolean{articletitles}}{\emph{{Performance of the \lhcb RICH
  detector at the LHC}},
  }{}\href{https://doi.org/10.1140/epjc/s10052-013-2431-9}{Eur.\ Phys.\ J.\
  \textbf{C73} (2013) 2431},
  \href{http://arxiv.org/abs/1211.6759}{{\normalfont\ttfamily
  arXiv:1211.6759}}\relax
\mciteBstWouldAddEndPuncttrue
\mciteSetBstMidEndSepPunct{\mcitedefaultmidpunct}
{\mcitedefaultendpunct}{\mcitedefaultseppunct}\relax
\EndOfBibitem
\bibitem{LHCb-DP-2013-001}
F.~Archilli {\em et~al.},
  \ifthenelse{\boolean{articletitles}}{\emph{{Performance of the muon
  identification at LHCb}},
  }{}\href{https://doi.org/10.1088/1748-0221/8/10/P10020}{JINST \textbf{8}
  (2013) P10020}, \href{http://arxiv.org/abs/1306.0249}{{\normalfont\ttfamily
  arXiv:1306.0249}}\relax
\mciteBstWouldAddEndPuncttrue
\mciteSetBstMidEndSepPunct{\mcitedefaultmidpunct}
{\mcitedefaultendpunct}{\mcitedefaultseppunct}\relax
\EndOfBibitem
\bibitem{LHCb-DP-2012-004}
R.~Aaij {\em et~al.}, \ifthenelse{\boolean{articletitles}}{\emph{{The \lhcb
  trigger and its performance in 2011}},
  }{}\href{https://doi.org/10.1088/1748-0221/8/04/P04022}{JINST \textbf{8}
  (2013) P04022}, \href{http://arxiv.org/abs/1211.3055}{{\normalfont\ttfamily
  arXiv:1211.3055}}\relax
\mciteBstWouldAddEndPuncttrue
\mciteSetBstMidEndSepPunct{\mcitedefaultmidpunct}
{\mcitedefaultendpunct}{\mcitedefaultseppunct}\relax
\EndOfBibitem
\bibitem{Sjostrand:2006za}
T.~Sj\"{o}strand, S.~Mrenna, and P.~Skands,
  \ifthenelse{\boolean{articletitles}}{\emph{{PYTHIA 6.4 physics and manual}},
  }{}\href{https://doi.org/10.1088/1126-6708/2006/05/026}{JHEP \textbf{05}
  (2006) 026}, \href{http://arxiv.org/abs/hep-ph/0603175}{{\normalfont\ttfamily
  arXiv:hep-ph/0603175}}\relax
\mciteBstWouldAddEndPuncttrue
\mciteSetBstMidEndSepPunct{\mcitedefaultmidpunct}
{\mcitedefaultendpunct}{\mcitedefaultseppunct}\relax
\EndOfBibitem
\bibitem{Sjostrand:2007gs}
T.~Sj\"{o}strand, S.~Mrenna, and P.~Skands,
  \ifthenelse{\boolean{articletitles}}{\emph{{A brief introduction to PYTHIA
  8.1}}, }{}\href{https://doi.org/10.1016/j.cpc.2008.01.036}{Comput.\ Phys.\
  Commun.\  \textbf{178} (2008) 852},
  \href{http://arxiv.org/abs/0710.3820}{{\normalfont\ttfamily
  arXiv:0710.3820}}\relax
\mciteBstWouldAddEndPuncttrue
\mciteSetBstMidEndSepPunct{\mcitedefaultmidpunct}
{\mcitedefaultendpunct}{\mcitedefaultseppunct}\relax
\EndOfBibitem
\bibitem{LHCb-PROC-2010-056}
I.~Belyaev {\em et~al.}, \ifthenelse{\boolean{articletitles}}{\emph{{Handling
  of the generation of primary events in Gauss, the LHCb simulation
  framework}}, }{}\href{https://doi.org/10.1088/1742-6596/331/3/032047}{{J.\
  Phys.\ Conf.\ Ser.\ } \textbf{331} (2011) 032047}\relax
\mciteBstWouldAddEndPuncttrue
\mciteSetBstMidEndSepPunct{\mcitedefaultmidpunct}
{\mcitedefaultendpunct}{\mcitedefaultseppunct}\relax
\EndOfBibitem
\bibitem{Lange:2001uf}
D.~J. Lange, \ifthenelse{\boolean{articletitles}}{\emph{{The EvtGen particle
  decay simulation package}},
  }{}\href{https://doi.org/10.1016/S0168-9002(01)00089-4}{Nucl.\ Instrum.\
  Meth.\  \textbf{A462} (2001) 152}\relax
\mciteBstWouldAddEndPuncttrue
\mciteSetBstMidEndSepPunct{\mcitedefaultmidpunct}
{\mcitedefaultendpunct}{\mcitedefaultseppunct}\relax
\EndOfBibitem
\bibitem{Golonka:2005pn}
P.~Golonka and Z.~Was, \ifthenelse{\boolean{articletitles}}{\emph{{PHOTOS Monte
  Carlo: A precision tool for QED corrections in $Z$ and $W$ decays}},
  }{}\href{https://doi.org/10.1140/epjc/s2005-02396-4}{Eur.\ Phys.\ J.\
  \textbf{C45} (2006) 97},
  \href{http://arxiv.org/abs/hep-ph/0506026}{{\normalfont\ttfamily
  arXiv:hep-ph/0506026}}\relax
\mciteBstWouldAddEndPuncttrue
\mciteSetBstMidEndSepPunct{\mcitedefaultmidpunct}
{\mcitedefaultendpunct}{\mcitedefaultseppunct}\relax
\EndOfBibitem
\bibitem{Allison:2006ve}
Geant4 collaboration, J.~Allison {\em et~al.},
  \ifthenelse{\boolean{articletitles}}{\emph{{Geant4 developments and
  applications}}, }{}\href{https://doi.org/10.1109/TNS.2006.869826}{IEEE
  Trans.\ Nucl.\ Sci.\  \textbf{53} (2006) 270}\relax
\mciteBstWouldAddEndPuncttrue
\mciteSetBstMidEndSepPunct{\mcitedefaultmidpunct}
{\mcitedefaultendpunct}{\mcitedefaultseppunct}\relax
\EndOfBibitem
\bibitem{Agostinelli:2002hh}
Geant4 collaboration, S.~Agostinelli {\em et~al.},
  \ifthenelse{\boolean{articletitles}}{\emph{{Geant4: A simulation toolkit}},
  }{}\href{https://doi.org/10.1016/S0168-9002(03)01368-8}{Nucl.\ Instrum.\
  Meth.\  \textbf{A506} (2003) 250}\relax
\mciteBstWouldAddEndPuncttrue
\mciteSetBstMidEndSepPunct{\mcitedefaultmidpunct}
{\mcitedefaultendpunct}{\mcitedefaultseppunct}\relax
\EndOfBibitem
\bibitem{LHCb-PROC-2011-006}
M.~Clemencic {\em et~al.}, \ifthenelse{\boolean{articletitles}}{\emph{{The
  \lhcb simulation application, Gauss: Design, evolution and experience}},
  }{}\href{https://doi.org/10.1088/1742-6596/331/3/032023}{{J.\ Phys.\ Conf.\
  Ser.\ } \textbf{331} (2011) 032023}\relax
\mciteBstWouldAddEndPuncttrue
\mciteSetBstMidEndSepPunct{\mcitedefaultmidpunct}
{\mcitedefaultendpunct}{\mcitedefaultseppunct}\relax
\EndOfBibitem
\bibitem{Breiman}
L.~Breiman, J.~H. Friedman, R.~A. Olshen, and C.~J. Stone, {\em Classification
  and regression trees}, Wadsworth international group, Belmont, California,
  USA, 1984\relax
\mciteBstWouldAddEndPuncttrue
\mciteSetBstMidEndSepPunct{\mcitedefaultmidpunct}
{\mcitedefaultendpunct}{\mcitedefaultseppunct}\relax
\EndOfBibitem
\bibitem{AdaBoost}
Y.~Freund and R.~E. Schapire, \ifthenelse{\boolean{articletitles}}{\emph{A
  decision-theoretic generalization of on-line learning and an application to
  boosting}, }{}\href{https://doi.org/10.1006/jcss.1997.1504}{J.\ Comput.\
  Syst.\ Sci.\  \textbf{55} (1997) 119}\relax
\mciteBstWouldAddEndPuncttrue
\mciteSetBstMidEndSepPunct{\mcitedefaultmidpunct}
{\mcitedefaultendpunct}{\mcitedefaultseppunct}\relax
\EndOfBibitem
\bibitem{Pivk:2004ty}
M.~Pivk and F.~R. Le~Diberder,
  \ifthenelse{\boolean{articletitles}}{\emph{{sPlot: A statistical tool to
  unfold data distributions}},
  }{}\href{https://doi.org/10.1016/j.nima.2005.08.106}{Nucl.\ Instrum.\ Meth.\
  \textbf{A555} (2005) 356},
  \href{http://arxiv.org/abs/physics/0402083}{{\normalfont\ttfamily
  arXiv:physics/0402083}}\relax
\mciteBstWouldAddEndPuncttrue
\mciteSetBstMidEndSepPunct{\mcitedefaultmidpunct}
{\mcitedefaultendpunct}{\mcitedefaultseppunct}\relax
\EndOfBibitem
\bibitem{LHCb-PAPER-2013-065}
LHCb collaboration, R.~Aaij {\em et~al.},
  \ifthenelse{\boolean{articletitles}}{\emph{{Measurements of the $\Bp$, $\Bz$,
  $\Bs$ meson and $\Lb$ baryon lifetimes}},
  }{}\href{https://doi.org/10.1007/JHEP04(2014)114}{JHEP \textbf{04} (2014)
  114}, \href{http://arxiv.org/abs/1402.2554}{{\normalfont\ttfamily
  arXiv:1402.2554}}\relax
\mciteBstWouldAddEndPuncttrue
\mciteSetBstMidEndSepPunct{\mcitedefaultmidpunct}
{\mcitedefaultendpunct}{\mcitedefaultseppunct}\relax
\EndOfBibitem
\end{mcitethebibliography}

\newpage


 
\newpage
\centerline{\large\bf LHCb collaboration}
\begin{flushleft}
\small
R.~Aaij$^{27}$,
B.~Adeva$^{41}$,
M.~Adinolfi$^{48}$,
C.A.~Aidala$^{74}$,
Z.~Ajaltouni$^{5}$,
S.~Akar$^{59}$,
P.~Albicocco$^{18}$,
J.~Albrecht$^{10}$,
F.~Alessio$^{42}$,
M.~Alexander$^{53}$
% ...

{\footnotesize \it
% ...
$ ^{5}$Clermont Universit{\'e}, Universit{\'e} Blaise Pascal, CNRS/IN2P3, LPC, Clermont-Ferrand, France\\
% ...
$ ^{10}$Fakult{\"a}t Physik, Technische Universit{\"a}t Dortmund, Dortmund, Germany\\
% ...
$ ^{18}$INFN Laboratori Nazionali di Frascati, Frascati, Italy\\
% ...
$ ^{27}$Nikhef National Institute for Subatomic Physics, Amsterdam, Netherlands\\
% ...
$ ^{41}$Instituto Galego de F{\'\i}sica de Altas Enerx{\'\i}as (IGFAE), Universidade de Santiago de Compostela, Santiago de Compostela, Spain\\
$ ^{42}$European Organization for Nuclear Research (CERN), Geneva, Switzerland\\
% ...
$ ^{48}$H.H. Wills Physics Laboratory, University of Bristol, Bristol, United Kingdom\\
% ...
$ ^{53}$School of Physics and Astronomy, University of Glasgow, Glasgow, United Kingdom\\
% ...
$ ^{59}$University of Cincinnati, Cincinnati, OH, United States\\
% ...
$ ^{61}$Syracuse University, Syracuse, NY, United States\\
% ...
$ ^{74}$University of Michigan, Ann Arbor, United States, associated to $^{61}$\\
% ...
}
\end{flushleft}

\end{document}